\documentclass[12 pt]{amsart}
\usepackage{url}
\usepackage{hyperref}

\usepackage{graphicx}
\usepackage{amsmath}
\usepackage{amssymb}
\usepackage{enumerate}
\usepackage{lineno}
\usepackage[center]{caption}
\usepackage[section]{algorithm}
\usepackage[noend]{algpseudocode}

\newcommand{\cH}{\mathcal{H}}

 \newtheorem{theorem}{Theorem}[section]
 
 \newtheorem{lemma}[theorem]{Lemma}
 
 \theoremstyle{definition}
 
 \theoremstyle{remark}
 \newtheorem{remark}[theorem]{Remark}
 
 \numberwithin{equation}{section}

\begin{document}
\title[Separation of quantum gates]{Separation of gates in quantum parallel programming}

\author{Kan He}
\address{College of Information and Computer \& College of Mathematics, Taiyuan
 University of Technology, Taiyuan, Shanxi,030024, P. R. China} \email{hekanquantum@163.com}

\author{Shusen Liu}
\address{Institute for Quantum Computing, Baidu Research, Beijing 100193, P. R. China}
\address{Institute for Advanced Study, Tsinghua University, Beijing 100084, P. R. China}
\email{shusen88.liu@gmail.com}

\author{Jinchuan Hou}
\address{College of Mathematics, Taiyuan University of Technology, Taiyuan, Shanxi,
030024, P. R. China}\email{jinchuanhou@aliyun.com}

\begin{abstract}

     The number of qubits in current quantum computers is a major restriction on their wider application.
    To address this issue, Ying conceived of using two or more small-capacity quantum computers to produce
    a larger-capacity quantum computing system by quantum parallel programming ([M. S. Ying,  Morgan-Kaufmann, 2016]).
    In doing so, the main obstacle
    is separating the quantum gates in the whole circuit to produce a tensor product of the local gates.
    In this study, we theoretically analyse the (sufficient and necessary) separability conditions of
    multipartite quantum gates in finite or infinite dimensional systems. We then conduct separation experiments
    with n-qubit quantum gates on IBM quantum computers using QSI software.

\end{abstract}

\maketitle

\vspace{3mm}
\section{Introduction}

With the development of quantum hardware, programming for quantum
computers has become an urgent task \cite{sof,qlanguage,mue,zeng}.
As reported in ~\cite{sof,sel,gay,yingbook}, extensive research has
been conducted on quantum programming over the last decade, and
several quantum programming platforms have been developed over the
last two decades. The first quantum programming environment was the
`QCL' project proposed by \"{O}mer in 1998~\cite{qcl1,qcl2}. In
2003, Bettelli et al. defined a quantum language called Q language
as a C++ library~\cite{qlanguage}. In recent years, more scalable
and robust quantum programming platforms have emerged. In 2013,
Green et al. proposed a scalable functional quantum programming
 language, called Quipper, using Haskell as the host language~\cite{green}. JavadiAbhari et al. defined Scafford in 2014~\cite{Scafford}, presenting its accompanying compilation
  system ScaffCC~\cite{scaffCC}. Wecker and Svore from QuArc (the Microsoft Research Quantum Architecture and Computation team) developed
   LIQU$i|\rangle$ as a modern tool-set embedded within F\#~\cite{Liquid}.
At the end of 2017, QuARC announced a new programming language and
simulator designed specifically for full-stack quantum computing,
known as Q\#, which represents a milestone in quantum programming.
 In the same year, Liu et al. released the quantum program Q$|SI\rangle$  that supports a more complicated loop structure~\cite{liu2017q}.
 To date, the structures of programming languages and tools have mainly been sequential.
 However, beyond the constraints of quantum hardware,
 there remain several barriers to the development of practical
  applications for quantum computers. One of the most serious
   barriers is the number of physical qubits provided in physical machines.
    For example, IBMQ produces two five-qubits quantum computers ~\cite{ibmqx2} and one
    16-qubit quantum computer ~\cite{ibmqx3}, which are available to programmers through the cloud,
     but these are far fewer qubits than are required by practical quantum algorithms.
Today, quantum hardware is in its infancy. As the number of
available qubits is gradually increasing, many researchers are
considering the possibility of combining various quantum hardware
components to work as a single entity and thereby enable advances in
the number of qubits~\cite{yingbook}. To increase the number of
accessible qubits in quantum hardware, one approach uses concurrent
or parallel quantum programming. Although current quantum-specific
environments are sequential in structure, some researchers are
working to exploit the possibility of parallel or concurrent quantum
programming on the general programming platform from different
respects. Vizzotto and Costa applied mutually exclusive access to
global variables to enable concurrent programming in
Haskell~\cite{vizzotto2005concurrent}. Yu and Ying
 studied the termination of concurrent
programs~\cite{yu2012reachability}. Researchers provide mathematics
tools for process algebras to describe their interaction,
communication and
synchronization~\cite{gay2005communicating,feng2007probabilistic,ying2005pi,jorrand2004toward}.
Recently, Ying and Li  defined and established operational
(denotational) semantics and a series of proof rules for ensuring
the correctness of parallel quantum programs\cite{Yingli}.
 Recall quantum gates are unitary operators on a system, which are
 fundamental and common ingredients of quantum circuits.
Naturally, when implementing parallel quantum programs, the first
challenge is to separate multipartite quantum gates into the tensor
products of local gates. If separation is possible, a potential
parallel execution is the natural result. Here, we provide the
sufficient and necessary conditions for the separability of
multipartite gates. It is showed that multipartite quantum gates
that can be separated simply seldom exist. We then conduct
separation experiments with n-qubit quantum gates on IBM quantum
computers using QSI software.

\section{Criteria for separation of quantum gates and IBMQ experiments}

In this section, let $\cH_k$ be a separable complex Hilbert space of
finite or infinite dimension, $1\leq k\leq n$, and $ \otimes_{k=1}^n
\cH_k$ the tensor product of $\cH_k$s.
 Denote by $ \mathcal B(\otimes_{k=1}^n \cH_k), \mathcal U(\otimes_{k=1}^n \cH_k)$ and $\mathcal B_s(\otimes_{k=1}^n \cH_k)$
 respectively the algebra of all bounded linear operators, the group of all unitary operators (quantum gates), and the space of all self-adjoint operators
 on the underline space $\otimes_{k=1}^n \cH_k$.

Let $U$ be a multipartite gate on the composite system
$\otimes_{k=1}^n \cH_k$. We call that $U$ is separable (local or
decomposable) if there exist quantum gates $U_k$ on $\cH_k$ such
that
\begin{equation}
U=\otimes_{k=1}^n U_k\,.
\end{equation}
Considering each a unitary $U\in \mathcal U(\otimes_{k=1}^n \cH_k)$,
if $U=\exp[it{\bf H}]$ with ${\bf H}\in \mathcal B_s(\otimes_{k=1}^n
\cH_k)$, whether or not can we obtain separability criteria of $U$
from the structure of ${\bf H}$? Here, we first put forward the
separation problem for multipartite gates in the arbitrary (finite
or infinite) dimensional systems as follows.

{\bf The Separation Problem:} {\it Consider the multipartite system
$\otimes_{k=1}^n \cH_k$. If $U=\exp[i{\bf H}]$ with ${\bf
H}=\sum_{i=1}^{N_H} A_i^{(1)}\otimes A_i^{(2)} \otimes ... \otimes
A_i^{(n)}$ for
 a multipartite unitary gate $U$, determine whether there exist unitary
operators $U_k$ on $\cH_k$  such that $U=\otimes_{k=1}^n U_k$.
Further, how does the structure of each $U_k$ depend on the
exponents of $A_k^{(j)}$, $i=1, 2, ..., n$?}

\begin{remark}
    \label{rem:2.1}
Note that generally speaking, in the decomposition of ${\bf
H}=\sum_{i=1}^{N_H} A_i^{(1)}\otimes A_i^{(2)}\otimes ... \otimes
A_i^{(n)}$ in the above problem, there are many selections of the
operator set $\{A_i^{(j)}\}_{i,j}$ (even $A_i^{(j)}$ exist that may
not be self-adjoint). However, by \cite{hou3}, for an arbitrary
(self-adjoint or non-self-adjoint) decomposition ${\bf
H}=\sum_{i=1}^{N_H} B_i^{(1)}\otimes B_i^{(2)} \otimes ... \otimes
B_i^{(n)}$, there exists a self-adjoint decomposition ${\bf
H}=\sum_{i=1}^{N_H} A_i^{(1)}\otimes A_i^{(2)} \otimes ... \otimes
A_i^{(n)}$ such that
\begin{equation*}
{\rm span}\{B_1^{(j)}, B_2^{(j)}, ..., B_n^{(j)}\}={\rm
span}\{A_1^{(j)}, A_2^{(j)}, ..., A_n^{(j)}\}\,.
\end{equation*}
So we always assume that ${\bf H}$ takes its self-adjoint
decomposition in the following.

\end{remark}

To answer the separation question,  we begin the discussion with a
simple case: the length $N_H$ is 1, i.e., ${\bf H}=A_1\otimes
A_2\otimes ... \otimes A_n$. Let us first deal with the case $n=2$.

\begin{theorem}
    \label{theo:2.1}
    \it Let $\cH_1\otimes \cH_2$ be a  bipartite system of any  dimension. For a quantum gate $U=\exp [i{\bf H}]\in \mathcal
U(\cH_1\otimes \cH_2)$ with ${\bf H}=A\otimes B$, the following
statements are equivalent:
    \begin{enumerate}[{\rm (I)}]
        \item There exist unitary operators $C, D$ such that $U=C\otimes D$;
        \item One of $A, B$ belongs to ${\mathbb R}I$.
    \end{enumerate}

     Furthermore, there exist real scalars $\alpha, \beta$ such that either $C=\exp[i(tA+\alpha I)], D=I$ if $B=tI$, or $D=\exp[i(sB+\beta I)], C=I$ if $A=sI$.
\end{theorem}

Before giving the proof of Theorem \ref{theo:2.1}, we recall the
following lemma concerning the separate vectors of operator
algebras. Let $\mathcal A$ be a C$^*$-algebra on a Hilbert space
$\cH$. A vector $|x_0\rangle\in \cH$ is called a separate vector of
$\mathcal A$ if, for any $T\in \mathcal A$,
$T(|x\rangle)=0\Rightarrow T=0$. The following lemma is necessary to
complete the proof of Theorem \ref{theo:2.1}.

\begin{lemma}
    \label{lem:2.2} \cite{Conway}
    \it Every Abel C$^*$-algebra has separate vectors.
\end{lemma}

{\bf Proof of Theorem \ref{theo:2.1}.} (II)$\Rightarrow$ (I) is
obvious.  We only need to check  (I) $\Rightarrow$ (II).

Assume I). Then, for any unit vectors $|x\rangle, |x^\prime \rangle$
in the first system and $|y\rangle, |y^\prime \rangle$ in the second
system, one has

\begin{equation}
\begin{split}
\label{eq:2.2}
U|xy\rangle\langle x^\prime y^\prime| &=\exp[iA\otimes B]|xy\rangle\langle x^\prime y^\prime|     \\
&=|xy\rangle\langle x^\prime y^\prime|+iA\otimes B |xy\rangle\langle
x^\prime y^\prime|\\
&+ i^2\frac{A^2\otimes
    B^2|xy\rangle\langle x^\prime y^\prime|}{2!}+ ... \\
&\quad+i^k \frac{A^k\otimes B^k|xy\rangle\langle x^\prime
y^\prime|}{k!} + ...
\end{split}
\end{equation}
and,
\begin{equation}
\label{eq:2.3} U|xy\rangle\langle x^\prime y^\prime|=C\otimes
D|xy\rangle\langle x^\prime y^\prime|.
\end{equation}
Connecting Eq.~\ref{eq:2.2} and \ref{eq:2.3} and taking a partial
trace of the second (first) system respectively, we obtain that
$$
\begin{array}{rl}
\langle y|D|y^\prime\rangle C |x\rangle\langle x^\prime |=& \langle
y|y^\prime\rangle  |x\rangle\langle x^\prime |+i\langle
y|B|y^\prime\rangle A |x\rangle\langle x^\prime |\\ & +i^2\langle
y|B^2|y^\prime\rangle \frac{A^2}{2!}|x\rangle\langle x^\prime | +...
\\ &+i^k \langle y|B^k|y^\prime\rangle \frac{A^k}{k!}|x\rangle\langle
x^\prime |  +...
\end{array}$$
and $$\begin{array}{rl} \langle x|C|x^\prime\rangle D
|y\rangle\langle y^\prime |=& \langle x|x^\prime\rangle
|y\rangle\langle y^\prime |+i\langle x|A|x^\prime\rangle B
|y\rangle\langle y^\prime |
\\ &+i^2\langle
x|A^2|x^\prime\rangle \frac{B^2}{2!}|y\rangle\langle y^\prime | +...
\\ &+i^k \langle x|A^k|x^\prime\rangle \frac{B^k}{k!}|y\rangle\langle
y^\prime | +...\ .
\end{array}$$
 Then it follows from the arbitrariness of  $|x^\prime\rangle$
and $|y^\prime\rangle$ that
\begin{equation}
\label{eq:2.4} \begin{array}{rl} &\langle y|D|y^\prime\rangle C
|x\rangle \\ =& \langle y|y^\prime\rangle I|x\rangle+i\langle
y|B|y^\prime\rangle A|x\rangle+i^2 \langle y|B^2|y^\prime\rangle
\frac{A^2}{2!}|x\rangle+ ... \\ & +i^k \langle y|B^k|y^\prime\rangle
\frac{A^k}{k!}|x\rangle +...
\end{array}
\end{equation}
and
\begin{equation}
\label{eq:2.5} \begin{array}{rl}  &\langle x|C|x^\prime\rangle D
|y\rangle \\ &= \langle x|x^\prime\rangle I |y\rangle +i\langle
x|A|x^\prime\rangle B|y\rangle +i^2\langle x|A^2|x^\prime\rangle
\frac{B^2}{2!}|y\rangle + ...\\ & +i^k \langle x|A^k|x^\prime\rangle
\frac{B^k}{k!}|y\rangle +...
\end{array}
\end{equation}

Now there are the three cases that we should deal with.

{\bf Case 1. } $B=tI$. In this case, by taking $y^\prime=y$ in
Eq.~\ref{eq:2.4}, we see that
\begin{equation*}\begin{array}{rl}
&\langle y|D|y^\prime\rangle C |x\rangle\\ & =  I |x\rangle+ iA
|x\rangle+i^2 t^2 \frac{A^2}{2!} |x\rangle+ ... +i^k t^k
\frac{A^k}{k!} |x\rangle +...\\ &=\exp[itA] |x\rangle\,\end{array}
\end{equation*}
holds for all $|x\rangle$. Note that $C$ and $\exp[itA]$ are
unitary, so  there exists some $\alpha \in\mathbb R$ such that
$C=\exp[i\alpha ] \exp[itA]=\exp[i(tA+\alpha I)]$. It follows that
$U=\exp[i(tA+\alpha I)]\otimes I$.

{\bf Case 2. } $A=sI$. Similar to Case 1, in this case we have
$D=\exp[i\beta ] \exp[isB]=\exp[i(sB+\beta I)]$ for some
$\beta\in\mathbb R$. It follows that $U=I\otimes \exp[i(sB+\beta
I)]$.

{\bf Case 3. } $A, B\notin {\mathbb R}I$. In this case, a
contradiction  will be induced, so that Case 3 does not happen.
Dividing the following proof to the two subcases,

{\bf Subcase 3.1.  } Both $A$ and $ B$ have two distinct
eigenvalues. It follows that there exist two
 real numbers $t_1, t_2$ with $t_1\not= t_2$ such that $A|x_1\rangle=t_1
|x_1\rangle$ and $A|x_2\rangle=t_2 |x_2\rangle$, and $s_1,s_2$ with
$s_1\neq s_2$ such that $B|y_1\rangle=s_1 |y_1\rangle$ and
$B|y_2\rangle=s_2 |y_2\rangle$. Taking $|x\rangle=|x^\prime\rangle
=|x_1\rangle$ and $|x\rangle=|x^\prime\rangle =|x_2\rangle$ in
Eq.~\ref{eq:2.5} respectively, and $|y\rangle=|y^\prime\rangle
=|y_1\rangle$ and $|y\rangle=|y^\prime\rangle =|y_2\rangle$ in
Eq.~\ref{eq:2.4} respectively, we have that
\begin{equation*}
\langle x_1|C|x_1\rangle D=\exp[t_1 B],\quad \langle
x_2|C|x_2\rangle D=\exp[t_2 B]\,,
\end{equation*}
and
\begin{equation*} \langle y_1|D|y_1\rangle C=\exp[s_1 A], \quad
\langle y_2|D|y_2\rangle C=\exp[s_2 A]\,.
\end{equation*}
It follows that
\begin{equation*}
\langle x_1|C|x_1\rangle=\frac{\exp[s_1t_1]}{\langle
y_1|D|y_1\rangle} \quad{\rm and}\quad \langle
x_2|C|x_2\rangle=\frac{\exp[s_1t_2]}{\langle y_1|D|y_1\rangle}\,.
\end{equation*}
So one  gets
\begin{equation*}
\frac{\langle y_1|D|y_1\rangle \exp[t_1
    B]}{\exp[s_1t_1]}=D=\frac{\langle y_1|D|y_1\rangle \exp[t_2
    B]}{\exp[s_1t_2]}.
\end{equation*}
Taking the inner product for $|y_2\rangle$ on both sides of the
above equation, we have
\begin{equation*}
\frac{ \exp[t_1
    s_2]}{\exp[t_1 s_1]}=\frac{ \exp[t_2 s_2]}{\exp[t_2s_1]}\,.
\end{equation*}
It follows that $\exp[t_1 s_2-t_1 s_1]=\exp[t_2 s_2-t_2 s_1]$, which
leads to $t_1=t_2$ as $s_1-s_2\not=0$. This is a contradiction.

{\bf Subcase 3.2. } At least one of $A$ and $B$ has no  distinct
eigenvalues.

In this case, we must have
 dim$\cH_1\otimes \cH_2=\infty$ and at least one of $\sigma(A)$ and $\sigma (B)$, respectively the spectrum of $A$ and $B$, is an infinite closed subset of $\mathbb R$.
 With no loss of generality, say $\sigma(A)$  has infinite many points.   Let
$\mathcal A={\rm cl\ span}\{I, A, A^2, ..., A^n, ...\}$, then
$\mathcal A$ is a Abelian C$^*$-algebra. By Lemma \ref{lem:2.2},
$\mathcal A$ has a separate vector $|x_0\rangle$. Replacing
$|x\rangle$ with $|x_0\rangle$ and taking vectors the $|y\rangle,
|y'\rangle$ satisfying $\langle y|D|y'\rangle=0$ in
Eq.~\ref{eq:2.4}, we see that
\begin{equation}
\label{eq:2.6} \begin{array}{rl} 0=&\langle y|D|y^\prime\rangle C |x_0\rangle \\
= &\langle y|y^\prime\rangle I|x_0\rangle+\langle
y|B|y^\prime\rangle A|x_0\rangle- \langle y|B^2|y^\prime\rangle
\frac{A^2}{2!}|x_0\rangle- ...\\ & +i^k \langle
y|B^k|y^\prime\rangle \frac{A^k}{k!}|x_0\rangle +... \\= & (\sum_k
\lambda_k A^k) |x_0\rangle,
\end{array}
\end{equation}
where $\lambda_k=\frac{i^k \langle y|B^k|y^\prime\rangle}{k!}$. As
$|x_0\rangle$ is a separate vector, we must have $\sum_k \lambda_k
A^k=0$.

We claim that each $\lambda_k=0$. For any fixed $|y\rangle,
|y'\rangle$, note that the function $f(z)=\sum_k \lambda_k z^k$ is
analytic. Since $f(A)=0$, the spectrum  $\sigma(f(A))$ of $f(A)$
 contains the unique element 0. So, by the spectrum mapping theorem,
 we have
$$\{0\}=\sigma(f(A))=\{f(\lambda)| \lambda\in \sigma(A)\}.$$
Note that, by the assumption of this subcase, $\sigma(A)$ is an
infinite set and has at most one isolated point. So the analytic
function $f(z)$ must by zero. Then each $\lambda_k=0$. It follows
that, for each $  k=0,1,2, ...,n, ...$,
$$\langle y|B^k|y^\prime\rangle=0$$
holds for any vectors $|y\rangle, |y'\rangle$ satisfying $\langle
y|D|y'\rangle=0$. Particularly, for the case $k=0$, we have that,
for any vectors $|y\rangle, |y'\rangle$, $\langle
y|D|y'\rangle=0\Rightarrow \langle y|y^\prime\rangle=0$. This
ensures that $D\in {\Bbb R}I$. Now consider the case $k=1$, one
obtains that,  for any vectors $|y\rangle, |y'\rangle$, $\langle
y|D|y'\rangle=0\Rightarrow \langle y|B|y^\prime\rangle=0$. This
implies that $ B$ is linearly dependent to $D$. So we get $B\in
{\Bbb R}I$, which is a contradiction.

This completes the proof. \hfill$\square$

Next, we extend   Theorem \ref{theo:2.1} to the multipartite
systems. Before stating the result, let us give some notations.\\

Let $A_i$s be self-adjoint operators on ${\mathcal H}_i$, $i=1,2,
...,n$ such that ${\bf H}=A_1\otimes A_2 \otimes ... \otimes A_n$.
If there exists at most one element in the set $\{A_1, A_2, ... ,
A_n\}$ that does not belong to the set ${\mathbb R}I$, we can define
a scalar
\begin{equation}
\label{eq:2.6} \delta(A_j)=\begin{cases}
\prod_{k\neq j} \lambda_k,& \ {\rm if}\ A_j\notin {\mathbb R}I;  \\
0, &\ {\rm if} \ A_j\in {\mathbb R}I \\
\end{cases}
\end{equation}
where $A_k=\lambda_kI$ if $A_k\in {\mathbb R}I$.

Based on Theorem~\ref{theo:2.1}, we reach the following conclusion
in the multipartite case.

\begin{theorem}
    \label{theo:2.3}
    \it   Let $\otimes_{i=1}^n \cH_i$ be a  multipartite system of any dimension.  For a multipartite quantum gate $U=\exp[i{\bf H}]\in \mathcal U(\otimes_{i=1}^n \cH_i)$ with
    ${\bf   H}=A_1\otimes A_2 \otimes ... \otimes A_n$, the following statements are equivalent:
    \begin{enumerate}[{\rm (I)}]
        \item There exist unitary operators $C_i\in \mathcal U( \cH_i)$ $(i=1, 2, ..., n)$ such that
        $U=\otimes_{i=1}^n C_i$;
        \item At most one element in $\{A_i\}_{i=1}^n$   does not belong to ${\mathbb R}I$.
    \end{enumerate}

     Furthermore, there is a unit-model number $\lambda $ such that
       \begin{equation} \label{eq:2.6} U=\lambda \otimes_{j=1}^n \exp[i\delta(A_j) A_j],\end{equation}
       where  $\delta(A_j)$s are as that defined in Eq. 2.7.
\end{theorem}

{\bf Proof. } (II) $\Rightarrow$ (I) is straightforward. To prove
(I) $\Rightarrow$ (II), we use induction on $n$.

According to Theorem~\ref{theo:2.1}, (I) $\Rightarrow$ (II) is true
for $n=2$. Assume that the implication is true for $n=k$. Now let
$n=k+1$. We have that
\begin{equation*}\begin{array}{rl}
&\exp[iA_1\otimes A_2 \otimes ... \otimes A_{k+1}]=\exp[i{\bf
H}]\\=&C_1\otimes C_2 \otimes ... \otimes C_{k}\otimes C_{k+1}\\=&
T\otimes C_{k+1}\,. \end{array}\end{equation*} It follows from
Theorem~\ref{theo:2.1} that either $A_{k+1}\in {\mathbb R}I$ or
$A_1\otimes A_2 \otimes ... \otimes A_k\in {\mathbb R}I$. If
$A_1\otimes A_2 \otimes ... \otimes A_k\in {\mathbb R}I$, then each
$A_i$ belongs to ${\mathbb R}I$. According to the induction
assumption, (II) holds true. If $A_{k+1}\in {\mathbb R}I$, assume
that $A_{k+1}=wI$, then
$$\exp[i{\bf H}]=\exp[iwA_1\otimes A_2 \otimes ... \otimes
A_k]\otimes I=C_1\otimes C_2 \otimes ... \otimes C_{k}\otimes  I.$$
It follows from the induction assumption that (II) holds true. Eq.
(2.8) is obtained by repeating to use (II) in Theorem 2.2. We
complete the proof. \hfill$\square$

Next we deal with the general case of ${\bf H}$: $1<N_H<\infty$.
Assume that a multipartite quantum gate $U=\exp [-it {\bf H}]$ with
${\bf H}=\sum_{i=1}^{N_H}T_i$ and $T_i=A_i^{(1)}\otimes
A_i^{(2)}\otimes ... \otimes A_i^{(n)}$. If at most one element in
each set $\{A_i^{(1)}, A_i^{(2)}, ... , A_i^{(n)}\}$  does not
belong to the set ${\mathbb R}I$,  we define a function:
\begin{equation}
\label{eq:2.9} \delta(A_k^{(i)})=
\begin{cases}
\prod_{k\neq i} \lambda_j^{(k)}, & {\rm if}\  A_j^{(i)}\notin {\mathbb R}I ; \\
0, & {\rm if}\ A_j^{(i)}\in {\mathbb R}I,
\end{cases}
\end{equation}
where we denote $A_j^{(k)}=\lambda_j^{(k)}I$ if $A_j^{(k)}\in
{\mathbb R}I$. In the following theorem, we grasp a class of
separable multipartite gates.

Before the theorem, we recall  the Zassenhaus formula states that
\begin{equation}
\label{eq:2.7} \exp[A+B]=\exp[A]\exp[B]\mathcal P_z(A,B)\,,
\end{equation}
where $\mathcal P_z(A,B)=\Pi_{i=2}^\infty \exp[ C_i(A, B)]$ and each
 term $C_i(A,B)$ is a homogeneous Lie polynomial in variables $A,
B$, i.e., $C_i(A,B)$ is a linear combination (with rational
coefficients) of commutators of the form $[V_1 ... [V_2,... ,
[V_{m-1}, V_m] ... ]]$ with $V_i \in \{A, B\}$ \cite{Mag,Cas}.
Especially,  $C_2(A, B)=-\frac{1}{2} [A, B]$ and $C_3(A,
B)=\frac{1}{3} [B,[A, B]]+\frac{1}{6} [A,[A, B]]$. As it is seen, if
$\Pi_{i=2}^\infty \exp[ C_i(A, B)]$ is a multiple of the identity,
then $\exp[A]\exp[B]=\lambda \exp[A+B]$ for some scalar $\lambda$.
Particularly, if $AB=BA$, then $\Pi_{i=2}^\infty \exp[ C_i(A, B)]\in
{\mathbb C}I$. Furthermore, for the multi-variable case, we have
\begin{equation}
\label{eq:2.8}
\begin{split}
& \exp[\sum_{i=1}^N A_i] \\
=&\prod_{i=1}^N \exp[A_i]\mathcal P_z(A_{N-1},A_N)\\ & \cdot \mathcal P_z(A_{N-2}, A_{N-1}+A_N) ...\mathcal P_z(A_1, \sum_{j=2}^N A_j)\\
=&\prod_{i=1}^N\exp[A_i]\prod_{k=1}^N\mathcal P_z(A_k,
\sum_{j=k+1}^N A_j).
\end{split}
\end{equation}

\begin{theorem}
    \label{theo:2.5}
    \it  For a multipartite quantum gate $U\in\mathcal U(\otimes_{k=1}^n
    \cH_k)$,
    if $U=\exp [-it {\bf H}]$  with
${\bf H}=\sum_{i=1}^{N_H}T_i$ and $T_i=A_i^{(1)}\otimes
A_i^{(2)}\otimes ... \otimes A_i^{(n)}$ with $[T_k,T_l]=0$ for each
pair $k,l$,
    and   at most one element in each set $\{A_i^{(1)}, A_i^{(2)}, ... ,
    A_i^{(n)}\}$
   does  not belong to the set ${\mathbb R}I$, then up to a unit modular
   scalar,
    \begin{equation}
    \label{eq:2.10}
    U=  U^{(1)} \otimes U^{(2)} \otimes ... \otimes U^{(n)}\,,
    \end{equation}
    where $U^{(i)}$ is the local quantum gate on $H_i$,
    \begin{equation*}
    U^{(i)}=\prod_{k=1}^{N_H}\exp[it \delta(A_k^{(i)}) A_k^{(i)}]\,,
    \end{equation*}
    where $\delta_k^{(i)}$ is defined by Eq.~\ref{eq:2.9}.
\end{theorem}

{\bf Remark. }
    From Theorem~\ref{theo:2.5}, we can grasp a subclass of separable multipartite
    gates. Each element in the subclass is of the form $\{U=\exp [-it \sum_{i=1}^{N_H}T_i]\in\mathcal U(\otimes_{k=1}^n
    \cH_k)$ with $T_i=A_i^{(1)}\otimes A_i^{(2)}\otimes ... \otimes
A_i^{(n)}$ satisfying $ [T_k,T_l]=0$ for each a pair $\{k,l\}$,
 and at most one element in $\{A_i^{(1)}, A_i^{(2)}, ... ,
    A_i^{(n)}\}$
    does not belong to $ {\mathbb R}I\}$.

{\bf Proof of Theorem~\ref{theo:2.5} } Let us first observe that for
any real number $r$, $\exp[rT]=(\exp[T])^r$. Furthermore,
$\exp[rT\otimes S]=\exp[rT]\otimes \exp[rS]$ if $\exp[T\otimes
S]=\exp[T]\otimes \exp[S]$. Indeed, for arbitrary positive integer
$N$, it follows from Baker formula that $\exp[NT]=(\exp[T])^N$. In
addition, $\exp[T]=\exp[\frac{T}{M}\cdot M]$ gives
$\exp[\frac{T}{M}]=(\exp[T])^{\frac{1}{M}}$. So, for any rational
number $a$, we have $\exp[aT]=(\exp[T])^a$. As $\phi(a)=\exp[aT]$ is
continuous in $a\in [0,\infty)$ and $\exp[-T]=(\exp[T])^{-1}$, one
sees that $\exp[aT]=(\exp[T])^a$ holds for any real number $a$.

Now according to the assumption and the definition of
$\delta_j^{(i)}$,  write $\prod_{k=1}^N\mathcal P_z(T_k,
\sum_{j=k+1}^N T_j)=\lambda I$ since $[T_k,T_l] =0$ for each pair
$k,l$, it follows from Theorem~\ref{theo:2.3} and Eq.~\ref{eq:2.8}
that
\begin{equation*}
\begin{split}
 & U =\exp [it {\bf H}] = \exp [it (\sum_{i=1}^{N_H}  T_i)] \\ = &
\prod_{i=1}^{N_H} \exp [it T_i]\prod_{k=1}^N\mathcal P_z(T_k,
\sum_{j=k+1}^N T_j)\\ =&\lambda  \prod_{i=1}^{N_H} \exp [it T_i] \\=&\lambda \prod_{i=1}^{N_H} \exp [it A_i^{(1)}\otimes A_i^{(2)} \otimes ...\otimes A_i^{(n)}] \\
= &\lambda\prod_{k=1}^{N_H}\exp[it \delta_k^{(1)} A_k^{(1)}]\otimes
\prod_{k=1}^{N_H}\exp[it \delta_k^{(2)} A_k^{(2)}]\\ & \otimes ...
\otimes \prod_{k=1}^{N_H}\exp[it \delta_k^{(n)} A_k^{(n)}]\,.
\end{split}
\end{equation*}
Absorbing the unit modular scalar $\lambda$ and letting
$U^{(i)}=\prod_{k=1}^{N_H}\exp[it \delta_k^{(i)} A_k^{(i)}]$, we
complete the proof. \hfill$\square$

{\bf Example 2.6} Here, we show some simple separable two-qubit
gates. We assume that the Planck constant equals to one and denote
by $\sigma_X, \sigma_Y$ and $\sigma_Z$ the Pauli matrices
$\left(\begin{array}{cccccccccccccccccccccc}
0  & 1 \\
1       & 0\\
\end{array}\right)$, $\left(\begin{array}{cccccccccccccccccccccc}
0  & -i \\
i       & 0\\
\end{array}\right)$ and  $\left(\begin{array}{cccccccccccccccccccccc}
1  & 0 \\
0      & -1\\
\end{array}\right)$.

In the two-qubit composite spin-$\frac{1}{2}$ system, the total spin
operator $\mathbf S^2$ is defined by $\mathbf
S^2=S_X^2+S_Y^2+S_Z^2$, where $S_X=\sigma_X\otimes I +I\otimes
\sigma_X, $ $S_Y=\sigma_Y\otimes I +I\otimes \sigma_Y, $
$S_Z=\sigma_Z\otimes I +I\otimes \sigma_Z$ (\cite{Wheel}). The three
operators $S_X, S_Y, S_Z$ assign $X, Y, Z$ components of spin to the
composite system respectively. The $X$-spin quantum gate
$U_X=\exp[-it H_X]$ with the Hamiltonian $H_X=I\otimes
\sigma_X+\sigma_X\otimes I$. According to Theorem~\ref{theo:2.1},
$U_X$ is separable and
$$\begin{array}{lllll}
U_X=\exp[-it H_X]=\exp[-it \sigma_X]\otimes \exp[-it
\sigma_X].\end{array}$$ Similarly, $U_Y$ and $U_Z$ can be defined
analogously, and
$$\begin{array}{lllll} U_Y=\exp[-it H_Y]=\exp[-it \sigma_Y]\otimes
\exp[-it \sigma_Y].\end{array}$$
$$\begin{array}{lllll}
U_Z=\exp[-it H_Z]=\exp[-it \sigma_Z]\otimes \exp[-it
\sigma_Z].\end{array}$$

Furthermore, let us consider the so-called special 7-parameter
Hamiltonian introduced in \cite{Wheel}, where $${\bf H}=\sum_{i=1}^4
(a_i \sigma_i\otimes I +I\otimes b_i\sigma_i),$$ with $a_0+b_0={\rm
tr}({\bf H})$ (so with seven not eight parameters). Rewrite $\mathbf
H=(\sum_{i=1}^4 a_i \sigma_i)\otimes I +I\otimes
(\sum_{i=1}^4b_i\sigma_i)$.

$$\begin{array}{lllll}
U&=\exp[-it {\bf H}]=\exp[-it(\sum_{i=1}^4 a_i \sigma_i)]\otimes
\exp[-it(\sum_{i=1}^4b_i\sigma_i)].
\end{array}$$ \hfill$\square$

In the following we devote to designing an algorithm to check
whether or not a multipartite gate is separable in $n$-qubit case
(see  Algorithm 2.1). We perform the  experiments on the IBM quantum
processor \emph{ibmqx4}, while generate the circuits by
Q$|SI\rangle$ (the key code segments can be obtained in
https://github.com/klinus9542).

\begin{algorithm}[!htp]
    \caption{Check whether a unitary is separable or not}
    \label{alg:1}
    \begin{algorithmic}[1]
        \Require $U$
        \Ensure $Status$, $NonIndentiIndex$
        \Function {[Status, NonIndentiIndex]=CheckSeper}{$U$}
        \Comment{If separable, it can tell the status; otherwise it will answer nothing about the status}
        \State $H$ $\gets$ Hermitian value of $U$
        \For{index=1:Number of System}
        \If{PosChecker($H$,$index$)}
        \State \Return Status$\gets$\emph{Separable}
        \State \Return NonIndentiIndex$\gets$\emph{index}
        \EndIf
        \EndFor
        \EndFunction
        \Function{Status=PosChecker} {$H$,$index$}
        \Comment{Recurse solve this problem}
        \If{index == 1}
        \State Status$\gets$CheckPosLastDimN($H$)
        \Else
        \State
        \[
        \begin{pmatrix}
        C_{11}&C_{12}\\
        C_{21}&C_{22}
        \end{pmatrix}=H
        \] where $dim(C_{11})=dim(C_{12})=dim(C_{21})=dim(C_{22})=\frac{1}{2}*dim(H)$
        \If{$C_{12}$ and $C_{21}$ is \emph{NOT} all $0$ matrix}
        \State Status$\gets$0 \Comment {Counter-diagonal matrix is all $0$}
        \ElsIf {$C_{11}$ is \emph{NOT} equal to $C_{22}$}
        \State Status$\gets$0 \Comment {Ensure $C_{11}$ is a repeat of $C_{22}$}
        \Else
        \If {PosChecker($C_{11}$,$index-1$)}   \Comment{Recursion process sub-matrix}
        \State \Return Status $\gets$ $0$
        \Else
        \State \Return Status $\gets$ $1$
        \EndIf
        \EndIf
        \EndIf
        \EndFunction
        \\
        \Function{Status=CheckPosLastDimN}{$H$}
        \Comment{If the dimension of input matrix great or equal to $4$, conduct this process; otherwise return \emph{true}}
        \State
        \[
        \begin{pmatrix}
        C_{11}&C_{12}\\
        C_{21}&C_{22}
        \end{pmatrix}=H
        \] where $dim(C_{11})=dim(C_{12})=dim(C_{21})=dim(C_{22})=\frac{1}{2}*dim(H)$
        \If {$C_{11}$, $C_{12}$, $C_{21}$ and $C_{22}$ are all diagonal matrix with only 1 element }
        \State Status $\gets$ 0;
        \Else
        \State Status $\gets$ 1;  \EndIf  \EndFunction
    \end{algorithmic}
\end{algorithm}

\section{Conclusion and discussion}

We established a number of evaluation criteria for  the separability
of multipartite gates. These criteria demonstrate that almost all
$A\in \{A_i\}^n_{i=1}$ should belong to $\mathbb{R}I$ for a
separable multipartite gate $U=\exp[i\mathbf{H}]$, where
$\mathbf{H}=A_1 \otimes A_2 \otimes \ldots \otimes A_n$. Most of
random multipartite gates cannot fundamentally satisfy the
separability condition in Theorem~\ref{theo:2.3}. So we will put
forward and discuss preliminarily the approximate separation
question, which has the more practical meaning. Roughly speaking,
the multipartite unitary  is closed to some local unitary when their
Hamiltonians are close to each other.

This work reveals that there are very few quantum computational tasks
(quantum circuits) that can be automatically parallelized.
Concurrent quantum programming and parallel quantum programming
still needs to be researched for a greater understanding of quantum
specific features concerning the separability of quantum states, local
operations and classical communication and even quantum networks.

{\bf Acknowledgements} Thanks for comments. Correspondence should be
addressed to Kan He (email: hekanquantum@163.com),  Shusen Liu
(email: shusen88.liu@gmail.com) and Jinchuan Hou (email:
jinchuanhou@aliyun.com). Shusen Liu contributed equally to Kan He.
The work is supported by National Natural Science Foundation of
China under Grant No. 11771011, 11671294, 61672007.

\end{document}